\documentclass[12pt]{article}

\newcommand{\la}{\lambda}

\newcommand{\Om}{\Omega}
\newcommand{\om}{\omega}

\newcommand{\beq}{\begin{equation}}
\newcommand{\eeq}{\end{equation}}

\usepackage{latexsym,amsfonts,amsmath,amsthm,amssymb}

\usepackage[T1]{fontenc}
\usepackage[latin1]{inputenc}
\title{On the representation of contextual probabilistic dynamics
in the complex Hilbert space: linear and nonlinear evolutions,
Schr\"odinger dynamics}
\author{Andrei Khrennikov\footnote{International Center for Mathematical Modeling
in Physics and Cognitive Sciences, Email:
Andrei.Khrennikov@msi.vxu.se; supported by EU-Network
 "QP and Applications'' and Nat. Sc. Found., grant N PHY99-07949 at KITP, Santa-Barbara. } \\
MSI, University of V\"axj\"o, S-35195, Sweden}

\date{}
\begin{document}
\maketitle

\abstract{We constructed the representation of contextual
probabilistic dynamics in the complex Hilbert space. Thus dynamics
of the wave function can be considered as Hilbert space projections
of realistic dynamics in a ``prespace''. The basic condition for
representing of the prespace-dynamics is the law of statistical
conservation of energy -- conservation of probabilities.
Construction of the dynamical representation is an important step in
the development of contextual statistical viewpoint to quantum
processes. But the contextual statistical model is essentially more
general than the quantum one. Therefore in general the Hilbert space
projection of the ``prespace'' dynamics can be nonlinear and even
irreversible (but it is always unitary). There were found conditions
of linearity and reversibility of the Hilbert space dynamical
projection. We also found conditions for the conventional
Schr\"odinger dynamics (including time-dependent Hamiltonians). We
remark that in general even the Schr\"odinger dynamics is based just
on the statistical conservation of energy; for individual systems
the law of conservation of energy can be violated (at least in our
theoretical model). }

\section{Introduction}

In a series of papers [1]-[3] there was demonstrated that by using a
so called {\it contextual approach} to classical  probabilities we
can construct a representation of the conventional
measure-theoretical probabilistic model in a complex
Hilbert.\footnote{Here context is a complex of physical  conditions.
The contextual approach is based on just one postulate: {\it all
probabilities depend on complexes of physical conditions.} It is
meaningless to speak about probability without to specify a
context.} This representation is based on an interference
generalization of the well known formula of total probability [4].
In quantum physics this generalization is known as the formula of
{\it interference of probabilities}. In the opposite to the
conventional quantum theory, we obtain interference of probabilities
in the classical (but contextual) probabilistic framework, i.e.,
without to appeal to the Hilbert space formalism. Starting with the
interference formula of total probability we represent some class of
contexts (so called trigonometric contexts) by probabilistic complex
amplitudes and the famous Born's rule takes place. In the abstract
form this representation  coincides with the conventional quantum
(Hilbert space) representation. Our representation is based on a
pair of realistic variables (Kolmogorovian random variables) $a$ and
$b;$ so called {\it reference observables.} These realistic
observables are naturally represented by self-adjoint operators
$\hat{a}, \hat{b}$ in the complex Hilbert space. It is amazing that
these operators corresponding to ordinary random variables do not
commute. Of course, the reference random variables should be chosen
in a special way. They should be {\it incompatible.} But
incompatibility of random variables is defined in classical
probabilistic (measure-theoretical) terms.

We remark that not all contexts can be represented by complex
probabilistic amplitudes. For example, there exist contexts inducing
so called {\it hyperbolic interference of probabilities.} Such
contexts are represented by probabilistic amplitudes which take
values in the two dimensional Clifford algebra (``algebra of
hyperbolic numbers''), see [1]--[3].

So in our approach there exists prequantum reality which can be
described by using the classical (contextual) probability theory. A
part of this reality can be represented in the quantum-like way (or
one can say: projected on the complex Hilbert space). There are many
ways to represent the prequantum probabilistic model in the complex
Hilbert space. Every representation is based on a pair of
observables. And that is the point! In opposition to common
considerations, see, e.g., J. von Neumann [5],  in our model only
two special observables are realistic. Therefore there are no
problems with no-go theorems: von Neumann, Kochen-Specker, Bell,...

In this paper on the basis of results of [1]-[3] we consider the
representation in the complex Hilbert space of realistic dynamics
for the reference variables. The basic assumption for the existence
of such a representation is the validity of the {\it law of
conservation of probabilities} for one of the reference observables,
e.g.,  $a.$ In particular, we can consider $a$ as the {\it energy
variable} and $b$ as the {\it position variable.} However, in
general our scheme of representation is more general and it can be
applied to all physical quantities with conservation of
probabilities. We emphasize that we do not assume conservation of
these quantities (e.g., the energy) for individual systems. In
particular, the law of conservation of energy can be violated for
some prequantum realistic dynamics (so we should distinguish the
laws of statistical conservation of energy and individual
conservation of energy).

Another unexpected feature of our model is that  dynamics in the
complex Hilbert space (representing prequantum realistic dynamics)
can be {\it nonlinear.} We found conditions of linearity of the
Hilbert space image of a prequantum dynamics. We remark that the
Hilbert space dynamics is always unitary (both in the linear and
nonlinear cases).

We also emphasize that in general the Hilbert space image of
prequantum dynamics can be {\it irreversible}. We found conditions of
reversibility. Finally, we found conditions which induce the
conventional Schr\"odinger dynamics: linear reversible unitary
dynamics. The Schr\"odinger dynamics is characterized
through dynamics of the interference coefficient (coefficient of statistical
disturbance) which appears in the interference formula of total
probability. Dynamics of this coefficient should be described by the
differential equation for {\it harmonic oscillator.}

\section{Contextual viewpoint to classical probability and interference of probabilities}

In this section we repeat the main points of contextual
measure-theoretical approach to interference of probabilities, see [1]--[3]  for details.

Let $(\Omega, {\cal F}, {\bf P})$ be a Kolmogorov probability
space.\footnote{This is a measure-theoretical model. Here $\Omega$
is an arbitrary set, ${\cal F}$ is a $\sigma$-field of subsets of
$\Omega$ and ${\bf P}$ is a normalized countably additive measure on
${\cal F}$ taking values in $[0,1]$ (Kolmogorov probability). Physical observables are represented  by random
variables  --  (measurable) functions on $\Omega.$ } By the standard
Kolmogorov axiomatics sets $A\in {\cal F}$ represent {\it{events.}}
In our simplest model of {\it{contextual probability}} (which can be
called the Kolmogorov contextual model) the same system of sets
${\cal F}$ is used to represent complexes of experimental physical
conditions -- {\it{contexts.}} The conditional probability is
mathematically defined by the Bayes' formula: ${\bf P}(A/C)={\bf
P}(AC)/{\bf P}(C), {\bf P}(C) \ne 0.$ In our contextual model this
probability has the meaning of the probability of occurrence of the
event $A$ under the complex of physical conditions $C.$

Let $a=a_1,...,a_n$ and $b=b_1,...,b_n$ be discrete random
variables. Then the classical {\it formula of total probability}
holds, see, e.g.,  [4]:
\begin{equation}
\label{DDK} {\bf P}(b=b_i/C)=\sum_n {\bf P}(a=a_n/C) {{\bf
P}(b=b_i/a=a_n, C)} \;.
\end{equation}
We remark that sets (belonging to ${\cal F}):$
\begin{equation}
\label{DD} B_x=\{\om \in \Om: b(\om)=x\}\; \;  \mbox{and} \; \;
A_y=\{\om \in \Om: a(\om)=y\}
\end{equation}
have two different interpretations. On the one hand, these sets
represent events corresponding to occurrence of the values $b=x$ and
$a=y,$ respectively. On the other hand, they represent contexts
(complexes of physical conditions) corresponding to selections of
physical systems with respect to values $b=x$ and $a=y,$
respectively. The main problem with the formula of total probability
is that in general it is impossible to construct a context ``$A_y
C$" corresponding to a selection with respect to the value $a=y$
which would not disturb systems prepared by the context $C:$ only in
the absence of disturbances induced by measurements we can use the
set theoretical operation of intersection. I would like to modify
the formula of total probability by eliminating from consideration
sets ``$A_y C$" which in general do not represent physically
realizable contexts.

 A set $C$ belonging to ${\cal F}$ is said to be
a {\it non degenerate context} with respect to the $a$-variable if
${\bf P}(A_y C)\not =0$ for all $y.$ We denote the set of such
contexts by the symbol ${\cal C}_{a, \rm{nd}}.$ Let $a, b$ be two
random variables. They are said to be {\it incompatible} if ${\bf
P}(B_x A_y) \not = 0$ for all their values $x$ and $y.$  Thus $a$
and $b$ are incompatible iff every $B_x$ is a non degenerate context
with respect to $a,$ $B_x \in {\cal C}_{a, \rm{nd}},$ and vice versa

We shall consider  the case of {\it incompatible dichotomous random variables}
$a=a_1, a_2, b=b_1, b_2.$ We set
$$
Y=\{a_1, a_2\}, X=\{b_1, b_2\}
$$
(``spectra'' of random variables $a$ and $b).$ In  [1]--[3]  there was
proved the following interference formula of total probability
(generalizing the classical formula of total probability).
$$
{\bf P}(b=x/C)=\sum_{j=1}^2 {\bf P}(a=a_j/C){\bf P}(B/a=a_j)
$$
$$
+ 2 \lambda(b=x/a,C)\sqrt{\prod_{j=1}^2{\bf P}(a=a_j/C) {\bf
P}(b=x/a=a_j)} ,
$$
where
$
\lambda(b=x/a, C)$
$$
= \frac{{\bf P}(/C) - \sum_{j=1}^{2} {\bf
P}(b=x/a=a_j){\bf P}(a=a_j/C)}{2\sqrt{{\bf P}(a=a_1/C) {\bf P}(
b=x/a_1){\bf P}(a=a_2/C) {\bf P}( b=x/a=a_2)}}
$$
To obtain this formula, we put the expression for $\lambda$ into the
sum and obtain identity. In fact, this formula is just a
representation of the probability ${\bf P}(b=x/C)$ in a special way.
The $\la(b=x/a,C)$ was called the {\it coefficients of statistical
disturbance}  [1]--[3] or the coefficient of incompatibility of
observables $a$ and $b.$
\medskip

Suppose that, for every $x \in X,$ $ \vert \lambda(b=x/a,
C)\vert\leq 1 \;.$ In this case we can introduce new statistical
parameters $\theta(b=x/a, C)\in [0,2 \pi]$ and represent the
coefficients of statistical disturbance in the trigonometric form:
$\la(b=x/a, C)=\cos \theta (b=x/a, C).$ Parameters $\theta(b=x/a,
C)$ are called  {\it{relative phases}} of the events $B_x=
\{\omega\in \Omega: b(\omega)=x\}$ with respect to the observable
$a$ (in the context $C$); or simply {\it probabilistic phases}. We
remark that in general there is no geometry behind these phases;
these are purely probabilistic parameters. By using the
trigonometric representation of the coefficients $\lambda$ we obtain
the well known  {\it formula of interference of probabilities} which
is typically derived by using the Hilbert space formalism.
\footnote{If both coefficients $\lambda$ are larger than one, we can
represent them as $\la(b=x/a, C)=\pm \cosh \theta (b=x/a, C)$ and
obtain the formula of hyperbolic interference of probabilities, see
[1]--[3]; there can also be found models with the  mixed
hyper-trigonometric behaviour, see [1]--[3].}

\section{Representation of the contextual classical probabilistic
model in the Hilbert space}

We recall that we consider  the case of incompatible dichotomous random
variables $a=a_1, a_2, b=b_1, b_2.$ This pair of variables will be
fixed. We call such variables {\bf reference variables.} For each
pair $a, b$ of reference variables we construct a representation of
the contextual Kolmogorov model in the Hilbert space (``quantum-like
representation''). In this paper we shall be interested only in the
representation of trigonometric contexts:
$$
{\cal C}^{\rm tr}=\{C\in{\cal C}_{\rm{a, nd}}:|\la(b=x/a, c)|\leq 1,
x\in X\}
$$
Of course, the system ${\cal C}^{\rm tr}$ depends on the choice of a
pair of reference observables, ${\cal C}^{\rm tr}\equiv {\cal
C}^{\rm tr}_{b/a}$ We set $p_C^a(y)={\bf P}(a=y/C), p_C^b(x)={\bf
P}(b=x/C), p(x/y)={\bf P}(b=x/a=y), x \in X, y \in Y.$
 Let $C\in {\cal C}^{\rm{tr}}.$ The interference formula of total probability can be
written in the following form:
\begin{equation}
\label{Two} p_c^b(x)=\sum_{y \in Y}p_C^a(y) p(x/y) + 2\cos
\theta_C(x)\sqrt{\Pi_{y \in Y}p_C^a(y) p(x/y)}\;,
\end{equation}
where $\theta_C(x)=\theta(b= x/a, C)= \pm \arccos \lambda(b=x/a, C),
x \in X.$ By using the elementary formula:
\begin{equation}
\label{Two1} D=A+B+2\sqrt{AB}\cos \theta=\vert \sqrt{A}+e^{i
\theta}\sqrt{B}|^2,
\end{equation}
for $A, B > 0, \theta\in [0,2 \pi],$ we can represent the
probability $p_C^b(x)$ as the square of the complex amplitude
(Born's rule):
\begin{equation}
\label{Born} p_C^b(x)=\vert\varphi_C(x)\vert^2 ,
\end{equation}
where a complex probability amplitude is defined (through
(\ref{Two}) and (\ref{Two1})) by
\begin{equation}
\label{EX1} \varphi(x)\equiv \varphi_C(x)=\sqrt{p_C^a(a_1)p(x/a_1)}
+ e^{i \theta_C(x)} \sqrt{p_C^a(a_2)p(x/a_2)} \;.
\end{equation}
We denote the space of functions: $\varphi: X\to {\bf C}$ by the
symbol $\Phi =\Phi(X, {\bf C}).$ Since $X= \{b_1, b_2 \},$ the $\Phi$ is
the two dimensional complex linear space. By using the
representation (\ref{EX1}) we construct the map
\begin{equation}
\label{MAP} J^{b/a}:{\cal C} \to \Phi(X, {\bf C})
\end{equation}
The $J^{b/a}$ maps contexts (complexes of, e.g., physical
conditions) into complex amplitudes. The representation
({\ref{Born}}) of probability is nothing other than the famous {\bf
Born rule.} The complex amplitude $\varphi_C(x)$ can be called a
{\bf wave function} of the complex of physical conditions, context
$C$  or a  (pure) {\it state.} \footnote{We underline that the
complex linear space representation (\ref{EX1}) of the set of
contexts ${\cal C}$ is based on a pair $(a,b)$ of incompatible
(Kolmogorovian) random variables. Here $\varphi_C=\varphi_C^{b/a}.$
We call random variables $a, b$ {\bf reference variables.}}
We set $e_x^b(\cdot)=\delta(x- \cdot).$ The Born's rule for complex
amplitudes (\ref{Born}) can be rewritten in the following form:
\begin{equation}
\label{BH} p_C^b(x)=\vert(\varphi_C, e_x^b)\vert^2 \;,
\end{equation}
where the scalar product in the space $\Phi(X, C)$ is defined by
the standard formula: $(\varphi, \psi) = \sum_{x\in X}
\varphi(x)\bar \psi(x).$ The system of functions $\{e_x^b\}_{x\in
X}$ is an orthonormal basis in the Hilbert space $H=(\Phi, (\cdot,
\cdot))$

Let $X \subset R.$ By using the Hilbert space representation
({\ref{BH}}) of the Born's rule  we obtain  the Hilbert space
representation of the expectation of the (Kolmogorovian) random
variable $b$:
\begin{equation}
\label{BI1} E (b/C)= \sum_{x\in X}x\vert\varphi_C(x)\vert^2=
\sum_{x\in X}x (\varphi_C, e_x^b) \overline{(\varphi_C, e_x^b)}=
(\hat b \varphi_C, \varphi_C) \;,
\end{equation}
where  the  (self-adjoint) operator $\hat b: E \to E$ is determined
by its eigenvectors: $\hat b e_x^b=x e^b_x, x\in X.$ This is the
multiplication operator in the space of complex functions
$\Phi(X,{\bf C}):$ $ \hat{b} \varphi(x) = x \varphi(x) $
 By (\ref{BI1}) the  conditional expectation of the Kolmogorovian
random variable $b$ is represented with the aid of the self-adjoint
operator $\hat b.$  Therefore it is natural to represent this random
variable (in the Hilbert space model)  by the operator $\hat b.$
We would like to have Born's rule not only for the $b$-variable, but
also for the $a$-variable:
\begin{equation}
\label{BBR} p_C^a(y)=\vert(\varphi, e_y^a)\vert^2 \;, y \in  Y .
\end{equation}
How can we define the basis $\{e_y^a\}$ corresponding to the
$a$-observable? Such a basis can be found starting with interference
of probabilities. We set $u_j^a=\sqrt{p_C^a(a_j)},
u_j^b=\sqrt{p_C^b(b_j)}, p_{ij}=p(b_j/a_i), u_{ij}=\sqrt{p_{ij}},
\theta_j=\theta_C(b_j).$ We have:
$$
\varphi_C=v_1^b e_1^b + v_2^b e_2^b, \;\mbox{where}\;\; v_j^b=u_1^a
u_{1j}  + u_2^a u_{2j} e^{i \theta_j}\;.
$$
Hence
\begin{equation}
\label{BI} p_C^b(b_j) =\vert v_j^b \vert^2 = \vert u_1^a u_{1j}  +
u_2^a u_{2j} e^{i \theta_j} \vert^2.
\end{equation}
This is the {\it interference representation of probabilities} that
is used, e.g., in quantum formalism.

For any context $C,$ we can represent the corresponding wave
function $\varphi=\varphi_{C}$ in the form: \begin{equation}
\label{0} \varphi=u_1^a e_1^a + u_2^a e_2^a,
\end{equation}
where
\begin{equation}
\label{Bas} e_1^a= (u_{11}, \; \; u_{12}) ,\; \; e_2^a= (e^{i
\theta_1} u_{21}, \; \; e^{i \theta_2} u_{22})
\end{equation}
We consider the {\it matrix of transition probabilities} ${\bf
P}^{b/a}=(p_{ij}).$ It is always a  {\it stochastic matrix:}
$p_{i1}+p_{i2}=1, i=1,2).$ We remind  that a matrix is called  {\it
double stochastic} if it is stochastic and, moreover, $p_{1j} +
p_{2j}=1, j=1,2.$ The  system $\{e_i^a\}$  of vectors corresponding
to the $a$-observable is an orthonormal basis (and so the Born's rule holds true, see
[3] for the details)  iff the matrix ${\bf
P}^{b/a}$ is double stochastic and probabilistic phases satisfy the
constraint:
\begin{equation}
\label{RT}
 \theta_2 - \theta_1= \pi \; \rm{mod} \; 2 \pi.
\end{equation}

 {\bf Theorem.} [3]  {\it We can construct the quantum-like
(Hilbert space) representation of a contextual Kolmogorov space such
that the Born's rule holds true for both reference variables iff the
matrix of transition probabilities ${\bf P}^{b/a}$ is double
stochastic.}

It will be always supposed that the ${\bf P}^{b/a}$ is double
stochastic.

In this case the $a$-observable is represented by the operator
$\hat{a}$ which is diagonal (with eigenvalues $a_i)$ in the basis
$\{e_i^a\}.$ The Kolmogorovian conditional average of the random
variable $a$ coincides with the quantum Hilbert space average:
$$
E(a/C)=\sum_{y \in Y} y p_C^a(y) = (\hat{a} \phi_C, \phi_C), \; C \in {\cal
C}^{\rm{tr}}.
$$

\section{Representation of contextual probabilistic dynamics in the
complex Hilbert space}

Let us assume that the reference observables $a$ and $b$ evolve with
time:
$$
a=a(t,\omega), \; \: b=b(t,\omega).
$$
To simplify considerations, we consider evolutions which do not
change ranges of values of the reference observables:
 $Y=\{a_1,a_2\}$ and $ X=\{b_1,b_2\}$ do not depend on time. Thus,
 for any $t,$ $a(t,\omega)\in Y$ and $b=b(t,\omega)\in X.$ These are
 random walks with two-points state spaces $Y$ and $X.$
Since our main aim is the contextual probabilistic realistic
reconstruction of QM, we should restrict our considerations to
evolutions with the trigonometric interference. We proceed under the
following assumption:

\medskip

(CTRB) (Conservation of trigonometric behavior) \; {\it The set of
trigonometric contexts does not depend on time: ${\cal
C}^{\rm{tr}}_{a(t)/b(t)} = {\cal C}^{\rm{tr}}_{a(t_0)/b(t_0)}.$}

By (CTB) if a context $C \in {\cal C}^{\rm{tr}}_{a(t_0)/b(t_0)},$
i.e., at the initial instant of time the coefficients of statistical
disturbance $\vert \lambda(b(t_0)=x/a(t_0), C)\vert \leq 1,$ then
the coefficients $\lambda(b(t)=x/a(t), C)$ will always fluctuate in
the segment $[0,1].$ \footnote{Of course, there can be considered
more general dynamics in which the trigonometric probabilistic
behaviour can be transformed into the hyperbolic one and vice versa.
But we shall not try to study the most general dynamics. Our aim is
to show that the conventional Schr\"odinger dynamics can be easily
found among contextual dynamics in the complex Hilbert space.}

 For each instant of time $t,$ we can use the formalism of
contextual quantization, see (\ref{EX1}): a context $C$ can be
represented by a complex probability amplitude:
$$
\varphi(t, x) \equiv \varphi_C^{b(t)/a(t)} (x)
$$
$$
=\sqrt{p_C^{a(t)} (a_1) p^{b(t)/a(t)} (x/a_1)} + e^{i
\theta_C^{b(t)/a(t)}(x)} \sqrt{p_C^{a(t)} (a_2) p^{b(t)/a(t)}
(x/a_2)} .
$$
We remark that the observable $a(t)$ is represented by the self-adjoint operator $\hat a(t)$ defined by its
with eigenvectors:
\[e_1^a(t)= \left( \begin{array}{ll}
\sqrt{p(t; b_1/a_1)}\\
\sqrt{p(t; b_2/a_1)}
\end{array}
\right ), \;\;\;
e_2^a(t)= e^{i \theta_C(t)} \left( \begin{array}{ll}
\sqrt{p(t; b_1/a_2)}\\
- \sqrt{p(t; b_2/a_1)}
\end{array}
\right ),
\]
where
$$
p(t; x/y)=p^{b(t)/a(t)}(x/y), \theta_C(t)=\theta_C^{b(t)/a(t)}(b_1)
$$
and where we set $e_j^a(t)\equiv e_j^{a(t)}.$ We recall that
$\theta_C^{b(t)/a(t)} (b_2) = \theta_C^{b(t)/a(t)}(b_1) + \pi,$
since the matrix of transition probabilities is assumed to be
double stochastic for all instances of time.

We shall describe dynamics of the wave function $\varphi(t, x)$
starting with following assumptions  (CP) and (CTP). Then these
assumptions will be completed by the set (a)-(b) of mathematical
assumptions which will imply the conventional Schr\"odinger
evolution.

\medskip

(CP)(Conservation of $a$-probabilities)\;{\it The probability
distribution of the $a$-observable is preserved in process of
evolution:} \beq \label{PC} p_C^{a(t)}(y)=p_C^{a(t_0)}(y), y \in Y, \eeq
for any context $C \in {\cal C}^{\rm{tr}}_{a(t_0)/b(t_0)}.$ This
statistical conservation of the $a$-quantity will have very
important dynamical consequences.  We also
assume that the law of  conservation of transition probabilities
holds:

\medskip

(CTP) (Conservation of transition probabilities)\;{\it Probabilities
$p(t; x/y)$ are conserved in the process of evolution:} \beq
\label{PCT}
 p(t; x/y)=p(t_0; x/y)\equiv p(x/y).
\eeq

\medskip

Under the latter assumption we have: \beq \label{BB} e_1^a(t) \equiv
e_1^a(t_0), e_2^a(t)=e^{i[\theta_C(t) - \theta_C(t_0)]} e_2^a(t_0).
\eeq

{\bf Remark 4.1.} {\small If the $a(t)$-basis evolves according to
(\ref{BB}), then $\hat{a}(t)=\hat{a}(t_0)=\hat{a}.$ Hence the whole
stochastic process $a(t,\omega)$ is represented by one fixed
self-adjoint operator. We emphasize that random variables
$a(t_1,\omega)$ and $a(t_2,\omega),$ $t_1 \not= t_2,$ can differ
essentially as functions of the random parameter $\omega.$
Nevertheless, they are represented by the same quantum operator.}

Thus under assumptions (CTRB), (CP) and (CTP) we have:
$$
\varphi(t)=u_1^a e_1^a(t) + u_2^a  e_2^a(t)=u_1^a e_1^a(t_0) + e^{i
\xi_C(t,t_0)} u_2^a e_2^a(t_0) ,
$$
where $u_j^a=\sqrt{p_C^{a(t_0)} (a_j)}, j=1,2,$ and
$$
\xi_C(t,t_0)= \theta_C(t) - \theta_C(t_0).
$$
Let us consider the unitary operator $\hat{U}(t, t_0): {\cal H} \to
{\cal H}$ defined by this transformation of basis: $e^a(t_0)\to
e^a(t).$ In the basis $e^a(t_0)=\{e_1^a(t_0), e_2^a(t_0)\}$ the
$\hat{U}(t, t_0)$ can be represented by the matrix:
\[\hat{U}(t, t_0)= \left( \begin{array}{ll}
1&0\\
0 & e^{i\xi_C(t,t_0)}
\end{array}
\right ).
\]
We obtained the following dynamics in the Hilbert space ${\cal H}$:
\beq \label{E} \varphi(t)=\hat{U}(t, t_0) \varphi(t_0) \eeq This
dynamics looks very similar to the Schr\"odinger dynamics in the
Hilbert space. However, the dynamics (\ref {E}) is essentially more
general than Schr\"odinger's dynamics. In fact, the unitary operator
$\hat{U}(t, t_0)=\hat{U}(t, t_0, C)$ depends on the context $C,$
i.e., on the initial state $\varphi(t_0): \hat{U}(t, t_0) \equiv
\hat{U}(t, t_0, \varphi(t_0)).$ So, in fact, we derived the
following dynamical equation: \beq \label{EBU} \varphi(t)=\hat{U}(t,
t_0, \varphi_0) \varphi_0,
\end{equation}
where, for any $\varphi_0, \hat{U}(t, t_0, \varphi_0)$ is a family of unitary operators.

The conditions (CTRB), (CP) and (CTP)  are  natural from the
physical viewpoint (if the $a$-observable is considered as an analog
of energy, see further considerations). But these conditions do not
imply that the Hilbert space image of the contextual  realistic
dynamics is a linear unitary dynamics.

\medskip

{\it In general the   Hilbert space projection of the realistic prequantum dynamics is
nonlinear.}

\medskip

 To obtain a linear dynamics, we should make the
following assumption:

\medskip

(CI) (Context independence of the increment of the probabilistic phase)
{\it The $\xi_C(t,t_0)=\theta_C(t)-\theta_C(t_0)$ does not depend on $C.$}

\medskip

Under this assumption the unitary operator $\hat{U}(t, t_0)$ does not depend on $C.$
\begin{equation}
\label{E1}
\hat{U}(t, t_0)= \left( \begin{array}{ll}
1&0\\
0 & e^{i \xi(t,t_0)}
\end{array}
\right ).
\end{equation}
Thus the equation (\ref {E}) is the equation of the linear unitary evolution. The main problem in these considerations is to find a physical basis of the condition (CI): the increment of statistical disturbance should be the same for all contexts, see section 5 for the detailed  analysis.
The linear unitary evolution (\ref{E})  is still essentially more
general than the conventional Schr\"odinger dynamics. To obtain the
Schr\"odinger evolution, we need  a few standard mathematical
assumptions:

\medskip

(a). Dynamics is continuous: the map $(t,t_0) \to \hat{U}(t,t_0)$ is continuous.\footnote{We recall that there is considered
the finite dimensional case. Thus there is no problem of the choice of topology.}

(b). Dynamics is deterministic.

(c). Dynamics is invariant with respect to time-shifts;  $\hat{U}(t,
t_0)$ depends only on  $t-t_0:$ $  \hat{U}(t,t_0)\equiv
\hat{U}(t-t_0).$

The assumption of determinism can be described by the following relation:
$$
\phi(t; t_0, \phi_0)=\phi(t; t_1, \phi(t_1; t_0, \phi_0)) , \; t_0
\leq t_1 \leq t,
$$
where $\phi(t; t_0, \phi_0)= \hat{U}(t,t_0) \phi_0.$

It is well known  that
under the assumptions (a), (b), (c)  the family of (linear) unitary operators $\hat{U}(t, t_0)$
corresponds to the one parametric group of unitary operators:
\begin{equation}
\label{EB}
\hat{U}(t)= e^{-\frac{i}{h} \hat{H}t},
\end{equation}
where $\hat{H}: {\cal H} \to {\cal H}$ is a self-adjoint operator.
Here $h > 0$ is a scaling factor (e.g., the Planck constant). We
have: \begin{equation} \label{EA}\hat{H} = \left(\begin{array}{ll}
0 & 0 \\
0 & E
\end{array}
\right),
\end{equation}
where
$$
E= - h \Big[\frac{\theta_C(t)-\theta_C(t_0)}{t-t_0} \Big].
$$
Hence the Schr\"odinger evolution in the complex Hilbert space
corresponds to the contextual probabilistic dynamics with the linear
evolution of the probabilistic phase:
\begin{equation}
\label{EAF} \theta_C(t) =\theta_C(t_0) - \frac{E}{h}(t-t_0).
\end{equation}
Let us consider a stochastic process (rescaling of the process
$a(t,\omega)):$
\begin{equation}
\label{TIT} H(t, \omega) = \left\{ \begin{array}{ll}
{0, \; \; a(t, \omega)=a_1}\\
{E, \; \; a(t, \omega)=a_2.}
\end{array}
\right .
\end{equation}
Since the probability distributions of the processes $a(t,\omega))$
and $H(t, \omega)$ coincide (up to rescaling  of ranges of values), we
have
\begin{equation}
\label{RS} p_C^{H(t)}(0)=p_C^{a(t)}(a_1)\equiv
p_C^{a(t_0)}(a_1)=p_C^{H(t_0)}(0)
\end{equation}
\begin{equation}
\label{RS1} p_C^{H(t)}(E)=p_C^{a(t)}(a_2)\equiv
p_C^{a(t_0)}(a_2)=p_C^{H(t_0)}(E).
\end{equation}
If $E > 0$ we can interpret $H(t,\omega)$ as the energy observable
and the operator $\hat{H}$ as its Hilbert space image. We emphasize,
see Remark 4.1, that the whole ``energy process'' $H(t,\omega)$ is
represented by a single self-adjoint nonnegative operator $\hat{H}$
in the Hilbert space. This operator, ``quantum Hamiltonian'', is the
Hilbert space projection of the energy process which is defined on
the ``prespace'' $\Omega,$ see [3]. In principle, random variables
$H(t_1,\omega), H(t_2,\omega), t_1 \not=t_2,$ can be very different
(as functions of $\omega).$ We have only the law of {\it statistical
conservation of energy:}
\begin{equation}
\label{RS2} p_C^{H(t)}(z)\equiv p_C^{H(t_0)}(z), \; z=0, E.
\end{equation}

In general (depending on dynamics of the coefficient of the
statistical disturbance) the eigenvalue $E$ need not be positive. So
in general we have a dynamical equation corresponding to some
statistically conserved quantity, see, e.g., [] for details. Of
course, the representation (\ref{EB}) is equivalent to the
Schr\"odinger equation:
$$
 ih \; \frac{d \hat{U}}{d t}(t) = \hat{H} \hat{U} (t), \; \; \hat{U} (0) =I,
$$
where $I$ is the unit operator.

\section{Characterization of Schr\"odinger dynamics through dynamics of the coefficients of statistical disturbance}

We discuss here consequences of the condition (CI) for the
measurable quantity, namely, the coefficient of statistical
disturbance:
$$
\lambda_C(t) \equiv \lambda(b(t)=b_1/a(t), C) =\cos \theta_C (t).
$$
By (CI) we have $\theta_C(t)- \theta_C(t_0)=q (t,t_0),$
where $q$ does not depend on $C.$ Thus $\theta_C^\prime(t)=q^\prime(t, t_0)\equiv f(t,t_0)$
does not depend on $C.$ We remark that in principle $f(t,t_0)$ can depend on $t_0,$
 see Remark 5.2 for details. Let us investigate an interesting special case:

\medskip

(DT) {\it The function $f$ does not depend on $t_0.$ }

\medskip

Here $f=f(t).$ Suppose that $f(t)$ is a continuous function. Hence
$\theta_C(t)= \theta_C(t_0) + \int_{t_0}^t f(s) d s$ and
$$
\lambda_C(t)= \cos [\theta_C(t_0) + \int_{t_0}^t f(s) d s].
$$
We have the following differential equation for the coefficient $\lambda_C(t):$
\begin{equation}
\label{EABB}
\lambda_C^\prime(t)= \pm f(t) \sqrt{1-\lambda_C^2(t)}.
\end{equation}
To find the coefficient of statistical disturbance (under above
assumptions), one should solve the Cauchy problem for the
differential equation (\ref{EABB})with the initial condition
\begin{equation}
\label{EABB1}
\lambda_C(t_0)= \lambda(b(t_0)=b_1/a(t_0), C) \equiv \cos \theta_C(t_0).
\end{equation}
In this case the assumption (CI) can be written in the form of the Cauchy problem
(\ref{EABB}), (\ref{EABB1}). The evolution family has the form:
\begin{equation}
\label{EB5} \hat{U}(t, t_0)= \left( \begin{array}{ll}
1&0\\
0 & e^{i \int_{t_0}^t f(s) d s}
\end{array}
\right ).
\end{equation}This evolution is continuous and deterministic. To prove that the condition of determinism
(b) holds, we  use the additivity of integral.
In general such evolutions are not invariant with respect to time-shifts;
they correspond to Schr\"odinger evolutions
with time dependent generators:
$$
i h \frac{d \hat{U}(t, t_0)}{d t} = \hat{H}(t)\hat{U}(t, t_0), \;\hat{U}(t_0, t_0)=I,
$$
where
\begin{equation}
\label{EC}\hat{H}(t)= \left( \begin{array}{ll}
0 & 0 \\
0 & E(t)
\end{array}
\right ),
\end{equation}
where $E(t) = - h f(t).$ Let us consider two very simple, but
illustrative examples.

{\bf Example 1.} (Schr\"odinger dynamics)
Let $f(t)= - E /h= \rm{Const}.$ Then $\theta_C(t)=\theta_0 - E(t-t_0)/h.$ Here
\[\hat{U}(t)= \left( \begin{array}{ll}
1 & 0\\
0 & e^{-i\frac{Et}{h}}
\end{array}
\right ).
\]
This is the Schr\"odinger dynamics; so here the conditions (a)-(c)
are automatically satisfied. We remark that in this case the
coefficient of statistical disturbance  satisfies the second order
differential equation, namely, the equation for {\it  harmonic
oscillations:}
\begin{equation}
\label{EC5}
\frac{d^2\lambda}{ dt^2}(t) + \omega^2 \lambda(t)=0,
\end{equation}
where $\omega = E/h.$ This is the direct consequence of the
representation $\lambda(t)= \cos[\theta_0 - E(t-t_0)/h].$ Hence:

\medskip

{\it The Schr\"odinger dynamics is characterized by the harmonic fluctuations of the
coefficient of statistical disturbance.}

\medskip

{\bf Example 2.} Let $f(t)=-E t/h, E >0.$ Here $\theta_C(t)=
\theta_C(t_0)- E(t^2-t_0^2)/2h$ and
\[\hat{U}(t, t_0)= \left( \begin{array}{ll}
1&0\\
0 & e^{ -i E(t^2-t_0^2)/2h}
\end{array}
\right ).
\]
This is a linear unitary  deterministic and continuous dynamics, but
it is not invariant with respect to  time-shifts. This is the Schr\"odinger dynamics
with time-dependent generator of evolution. The $\hat{H}(t)$ is
positively defined and it can be considered as time dependent
Hamiltonian.

{\bf Remark 5.1.} (Approximate reversibility of the Hilbert space
evolution) {\small If the function $f=f(t,t_0)$ depends nontrivially
on $t_0,$ then the evolution of the wave function is not
deterministic. It is {\it irreversible}. We remark that if
determinism of the evolution of the probabilistic phase is violated,
then $f(t,t_0)$ nontrivially depends on $t_0$ (and vice versa). The
violation of determinism for the phase evolution means that
$\theta_C(t)$ is not uniquely determined by $\theta_C(t_0);$ so the
following condition of determinism:
$$
\theta_C(t; \theta_C(t_1; \theta_C(t_0)))= \theta_C(t;  \theta_C(t_0))
$$
should be violated, $t_0 \leq t_1 \leq t.$ Here $\theta_C(t;
\theta_C(t_0))$ is the probabilistic phase at the instant of time
$t$ under the condition that it was equal to $\theta_C(t_0)$ at the
initial instant of time $t_0.$ We remark that reversible  Hilbert
space evolutions could appear as  approximations of irreversible
Hilbert space evolutions. Suppose that
$$
f(t,t_0) = f(t) + \epsilon f_1(t, t_0),
$$
where $\epsilon$ is negligibly small. Then by neglecting terms of
the $\epsilon$-magnitude we can approximately describe the Hilbert
space evolution as reversible. Of course, there should be chosen
some scale. It is natural to use the scale based on the Planck
constant $h.$ Thus if $\epsilon << h,$ then the irreversible
evolution
\begin{equation}
\label{EBN}
\hat{U}(t, t_0)= \left( \begin{array}{ll}
1&0\\
0 & e^{i [\int_{t_0}^t f(s) d s + \epsilon \int_{t_0}^t f_1(t, t_0) ds] }
\end{array}
\right ).
\end{equation}
can be approximately considered as the reversible evolution
(\ref{EB}).}

{\bf Remark 5.2.} (Approximate linearity of the Hilbert space evolution)
{\small Arguments which are similar to the arguments of  the
previous Remark can be applied to the problem of linearization of
general nonlinear dynamics in the complex Hilbert space. Let
$f=f(t)$ (so dynamics is deterministic) and let $f(t)$ be an
analytic function:
$$
f(t)=\sum_{n=0}^\infty f_n t^n \equiv f_0+ f_1(t).
$$
Suppose that
$$
\epsilon= \max_{0\leq t\leq T} \vert f_1(t) \vert << h.
$$ If $\epsilon << h,$ then the Hilbert space dynamics could be
approximately considered as a linear dynamics.}

We also can combine arguments of both Remarks.

\bigskip

{\bf Conclusion.} {\it A contextual realistic dynamics can be
represented (under assumptions (CTRB), (CP), and (CTP)) as a unitary
dynamics  in the complex Hilbert space. In general such a dynamics
is nonlinear and irreversible. Dynamics is linear iff the condition
(CI) holds. The contextual dynamics in the Hilbert space is reduced
to the  conventional Schr\"odinger evolution under the additional
assumptions (a)-(c).  In particular, the assumption (b) is implies
reversibily.  The Schr\"odinger dynamics is the Hilbert
space projection of the realistic   dynamics with harmonic
oscillations, see (\ref{EC5}), of the coefficient of statistical
disturbance. It might be that the reversible and linear
Schr\"odinger dynamics is just an approximation of an irreversible
and nonlinear dynamics in the Hilbert space.}

The author of this paper was strongly influenced by investigations
on various aspects of the conditional probabilistic approach to
quantum mechanics; especially important role was played by works
[6]--[16]; some elements of models presented in these works were
used in the process of creation of the present contextual
statistical model.

{\bf References}

1.  1. A. Yu. Khrennikov, {\it J. Phys.A: Math. Gen.} {\bf 34}, 9965
(2001);{\it J. Math. Phys.}{\bf 44}, 2471 (2003); {\it Phys. Lett.
A} {\bf 316}, 279 (2003);

2. A. Yu. Khrennikov,``On foundations of quantum theory,'' in {\it
Quantum Theory: Reconsideration of Foundations},  A. Yu. Khrennikov,
ed. (V\"axj\"o University Press, 2002), pp. 163-173.

3. A. Yu. Khrennikov,  {\it Annalen  der Physik} {\bf 12}, 575
(2003); ``On the classical limit for the hyperbolic quantum
mechanics,'' quant-ph/0401035; {\it J. Math. Phys.}, {\bf 45}, 902
(2004)

4. A. N. Shiryayev, {\it Probability} (Springer, New
York-Berlin-Heidelberg, 1984).

5. J. von Neumann, {\it Mathematical foundations of quantum
mechanics} (Princeton Univ. Press, Princeton, N.J., 1955).

6.  L. Accardi, Phys. Rep. {\bf 77}, 169(1981); ``The probabilistic
roots of the quantum mechanical paradoxes,'' in {\it The
wave--particle dualism:  A tribute to Louis de Broglie on his 90th
Birthday}, S. Diner, D. Fargue, G. Lochak, and F. Selleri, eds. (D.
Reidel Publ. Company, Dordrecht,  1984), pp. 47-55.

7. L. Accardi, {\it Urne e Camaleoni: Dialogo sulla realta, le leggi
del caso e la teoria quantistica} (Il Saggiatore, Rome, 1997).

8. L. Accardi, A. Fedullo, Lettere al Nuovo Cimento {\bf 34,}
161-172 (1982).

9.  L. E. Ballentine,  Rev. Mod. Phys. {\bf 42},  358 (1970); {\it
Quantum mechanics} (Englewood Cliffs, New Jersey, 1989);
``Interpretations of probability and quantum theory,'' Q. Prob.
White Noise Anal. {\bf  13},  71 (2001).

10.  L. E. Ballentine, {\it Quantum mechanics} (WSP,  Singapore,
1998).

11. W. M. De Muynck, ``Interpretations of quantum mechanics, and
interpretations of violations of Bell's inequality'', Q. Prob. White
Noise Anal. {\bf  13},  95 (2001).

12. W. M. De Muynck, {\it Foundations of quantum mechanics, an
empiricists approach} (Kluwer, Dordrecht, 2002).

13. S. P. Gudder, Trans. AMS {\bf 119},  428 (1965); {\it Axiomatic
quantum mechanics and generalized probability theory} (Academic
Press, New York, 1970).

14. S. P. Gudder, ``An approach to quantum probability,'' Quantum
Prob. White Noise Anal. {\bf  13}, 147 (2001).

15. A. Land\'e, {\it Foundations of quantum theory} (Yale Univ.
Press, 1955); {\it New foundations of quantum mechanics} (Cambridge
Univ. Press, Cambridge, 1968).

16.  G. W. Mackey, {\it Mathematical foundations of quantum
mechanics} (W. A. Benjamin INc, New York, 1963).

\end{document}